\documentclass[10pt]{article}

\usepackage{amsmath}

\usepackage{array}

\usepackage{appendix}

\usepackage{tocloft}                   

\usepackage{graphicx}

\usepackage{amsfonts}

\usepackage{amssymb}

\usepackage{mathrsfs}

\usepackage{yfonts}

\usepackage{euscript}

\usepackage{centernot}                 

\usepackage{ifsym}                     

\usepackage{lmodern}                   

\usepackage{upgreek}

\usepackage{mathtools}

\usepackage{color}

\usepackage{slantsc}
\usepackage{calligra}

\usepackage{bbold}          

\usepackage[T1]{fontenc}

\usepackage{epsf}

\usepackage{latexsym}

\usepackage{tipa}

\usepackage{makeidx}

\makeindex

\def\b{\begin{equation}}

\def\e{\begin{equation}}

\def\be{\begin{equation}}              

\def\ee{\end{equation}}

\def\beq{\begin{equation}}

\def\eeq{\end{equation}}

\def\bea{\begin{eqnarray}}

\def\eea{\end{eqnarray}}

\def\m{\mbox{ }}

\def\!{\hspace{-1.6667em}}

\def\c{\cite}

\def\n{\noindent}

\def\u{\underline}

\def\bLambda{\mbox{\boldmath$\Lambda$}}             

\def\biS{\mbox{\boldmath$S$}}

\def\sbiA{\mbox{\scriptsize\boldmath$A$}}

\def\sbia{\mbox{\scriptsize\boldmath$a$}}

\def\sbiO{\mbox{\scriptsize\boldmath$O$}}

\def\bia{\mbox{\boldmath$a$}}

\def\biA{\mbox{\boldmath$A$}}

\def\biQ{\mbox{\boldmath$Q$}}

\def\biS{\mbox{\boldmath$S$}}

\def\mH{\mbox{H}}

\def\bsigma{\mbox{\boldmath$\sigma$}}                   %

\def\bphi{\mbox{\boldmath$\phi$}}                   %

\def\sf{\mbox{\scriptsize f}}

\def\si{\mbox{\scriptsize i}}

\def\sn{\mbox{\scriptsize n}}

\def\sS{\mbox{\scriptsize S}}

\def\sbcC{\mbox{\boldmath \scriptsize ${\cal C}$}}

\def\sbcG{\mbox{\boldmath \scriptsize ${\cal G}$}}

\def\bscG{\mbox{{\boldmath \scriptsize${\cal G}$}}}

\def\sumi2{\sum\mbox{}_{\mbox{}_{\mbox{\scriptsize $i$=1}}}^2}

\def\sumi3{\sum\mbox{}_{\mbox{}_{\mbox{\scriptsize $i$=1}}}^3}

\def\sumABcycles3{\sum\mbox{}_{\mbox{}_{\mbox{\scriptsize cycles $A,B$=1}}}^{3}}

\def\sumCDcycles3{\sum\mbox{}_{\mbox{}_{\mbox{\scriptsize cycles $C,D$=1}}}^{3}}

\def\sumj3{\sum\mbox{}_{\mbox{}_{\mbox{\scriptsize $j$=1}}}^3}

\def\sumk3{\sum\mbox{}_{\mbox{}_{\mbox{\scriptsize $k$=1}}}^3}

\def\prodiA1{\prod\mbox{}_{\mbox{}_{\mbox{\scriptsize $i$=1}}}^{A - 1}}

\def\d{\textrm{d}}                                                  

\def\es{\m = \m}

\def\:={\m := \m}

\def\=:{\m =: \m}

\def\lFrg{\mbox{\Large$\mathfrak{g}$}}                         
\def\nFrg{\mbox{\large$\mathfrak{g}$}}                         
\def\Frg{\mbox{\normalsize $\mathfrak{g}$}}                    

\def\Hilb{\mbox{{\boldmath$\mathfrak{H}$}ilb}}                 

\def\scH{\mbox{\scriptsize ${\cal H}$}}                    

\def\scM{\mbox{\scriptsize ${\cal M}$}}                    

\def\FrQ{\mbox{\Large $\mathfrak{q}$}}                               
\def\Phase{\mbox{{\boldmath$\mathfrak{P}$}hase}}                     

\def\bFrR{\mbox{\boldmath$\mathfrak{R}$}}                            
\def\Rig-Phase{\bFrR\mbox{ig-}\Phase}                                
                                                                													   
\def\bFrR{\mbox{\boldmath$\mathfrak{R}$}}                            

\def\bFrR{\mbox{\boldmath$\mathfrak{R}$}}                            

\def\1mat{\u{\u{1}}}                                                 

\def\Positive-Modespace{\mbox{{\boldmath$\mathfrak{M}$}odespace$^+$}}

\def\POSITIVE-MODESPACE{\mbox{{\boldmath$\mathfrak{M}$}ODESPACE$^+$}}

\def\Kin-Hilb{\mbox{{\boldmath$\mathfrak{K}$}in-\Hilb}}                     

\def\Mid-Hilb{\mbox{{\boldmath$\mathfrak{M}$}id-\Hilb}}                     

\def\Dyn-Hilb{\mbox{{\boldmath$\mathfrak{D}$}yn-\Hilb}}                     

\def\5Star{\mbox{\Large$\star$}}              

\begin{document}

\begin{center}

\large{\bf Lie Theory suffices to understand, and Locally Resolve, the Problem of Time}\normalsize  

\vspace{.05in} 

{\bf Edward Anderson}

\end{center}

\begin{abstract}

The Lie claw digraph controls Background Independence and thus the Problem of Time and indeed the Fundamental Nature of Physical Law.    
This has been established in the realms of Flat and Differential Geometry with varying amounts of extra mathematical structure. 
This Lie claw digraph has Generator Closure at its centre (Lie brackets), 
Relationalism at its root (implemented by Lie derivatives), 
and, as its leaves, Assignment of Observables (zero commutants under Lie brackets) 
and Constructability from Less Structure Assumed (working if generator Deformation leads to Lie brackets algebraic Rigidity).					
This centre is enabled by automorphisms and powered by the Generalized Lie Algorithm extension of the Dirac Algorithm 
(itself sufficing for the canonical subcase, for which generators are constraints).  
The Problem of Time's facet ordering problem is resolved.  

\end{abstract}

$^1$ dr.e.anderson.maths.physics *at* protonmail.com

\vspace{-0.15in}

\section{Introduction}

Over 50 years ago, Wheeler, DeWitt and Dirac \cite{Battelle-DeWitt, Dirac} 
found a number of conceptual problems with combining General Relativity (GR) and Quantum Mechanics (QM).

Kucha\v{r} and Isham \cite{K92-I93} subsequently classified attempted resolutions of these problems; see \cite{APoT} for a summary.  
They moreover observed that attempting to extend one of these {\bf Problem of Time} facet's resolution to include a second facet 
has a strong tendency to interfere with the first resolution (nonlinearity).

The Author next identified \cite{APoT2, ABook} each facet's nature in a theory-independent manner.  
This is in the form of clashes between background-dependent           (including conventional QM) 
                                   and background-independent Physics (including GR).
[Background Dependence means a fixed background structure, while {\bf Background Independence} approximately means letting this structure be dynamical instead.
Other authors to date have usually restricted Background Independence to the metric and differential-geometric levels of structure.]  
See Sec 2 for the above and current Authors' names for the constituent facets. 
Working locally, the Author recently provided not only a resolution of each individual facet 
but also a consistent mathematical formulation for the joint treatment of all local facets \cite{ABook}. 
This resolves the above `facet interference mystery', as well as the facet ordering problem \c{K93}, i.e.\  in which order the facets are to be approached. 
Let us furthermore refer to each resolved Problem of Time facet as a successfully incorporated aspect of Background Independence. 
Let us finally clarify that all of the above could just as well be termed `{\bf Fundamental Nature of Physical Law}'. 
Despite terminology such as 'Problem of Time' or 'Background Independence' having been attached to this subject area, it is {\it highly} mainstream in significance.

The Author subsequently showed that \cite{ALRoPoT} this `Local Theory of Background Independence' Local Resolution of the Problem of Time's 
classical realization uses just \cite{XIV} Lie Theory \cite{Lie1890, Still, Lee2, Olver} (Secs 3 and 4).    
Since Lie Theory is very widely known and a staple of grad-school level Physics and continuum Mathematics, 
this is both a very major simplification and of pedagogical value. 
I.e.\ the current Article's collection of interlinked and hitherto unresolved Physics problems 
turns out to be resolved by applying a basic and widely-known area of Mathematics. 
Wider significance of this result is outlined in the Conclusion (Sec 5). 

\vspace{-0.15in}

\section{Facet and aspect names: from prior literature current true-names}

Let us begin (Fig 1) by giving previous literature's names for Wheeler, DeWitt and Dirac's Problem of Time facets, 
progressing to give clarifying `true names' to the Background Independence aspects underlying these.
This serves for readers to, firstly, spot longstanding problem names they are familiar with.
Secondly, to follow the arrows to find what `true names' this Article refers to these by.  
This also serves to simplify down to just four types of problem: {\bf Closure}, {\bf Relationalism}, {\bf Observables} and {\bf Constructability} [see Sec 3].

One reason that the Problem of Time has many facets and has hitherto caused confusion as to, firstly, what those facets are. 
Secondly, as to in which order they are to be addressed is that it contains {\sl two different} copies of the above quartet of problem. 
Namely the {\bf spacetime primality copy} and the {\bf space/dynamics/canonical primality copy}.  
Insufficient attention to, or presentational distinction between, these two copies contributed to confusion in the previous literature 
[see \cite{ABook, ALRoPoT} for details].  
The above quartet, moreover, constitutes {\bf Lie Theory}.  
One of the reasons Problem of Time facets interfere with each other 
turns out to be none other than the well-known fact that the parts of Lie Theory are inter-related.  
Facet ordering was also affected by confusion between copies, 
or failing to realize that many facets come in two distinct copies which can then occur at different places from each other in the facet ordering.
%
{            \begin{figure}[!ht]
\centering
\includegraphics[width=1.0\textwidth]{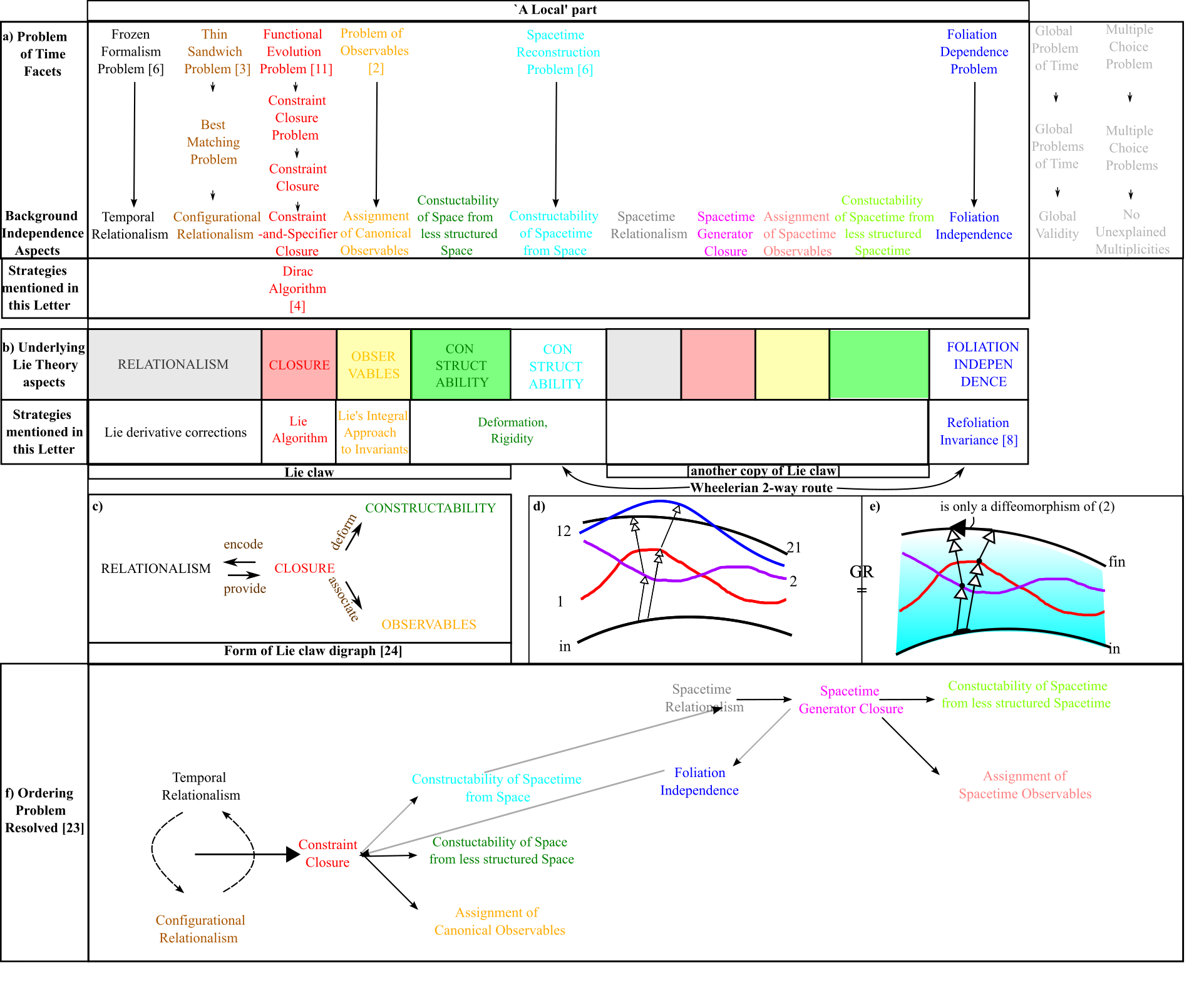}
\caption[Text der im Bilderverzeichnis auftaucht]{\footnotesize{a) Evolution of conceptualization and nomenclature of Problem of Time facets  
from Wheeler, DeWitt, Dirac, Kucha\v{r} and Isham 
to underlying Background Independence aspects including two copies [black arrows in e)] of the Lie claw digraph c). 
d) poses and resolves GR's Refoliation Invariance.
I.e.\ whether in going from an initial object to a final object, proceeding (= applying $\scH$) via the red surface 1 or the purple surface 2 
                                                                 causes discrepancy by at most a diffeomorphism of the thus-shared final surface, fin.} }
\label{Evol-Fac}\end{figure}            }

\vspace{-0.15in}
%
\section{The Lie claw digraph of facets of aspects}

Let us next address the generalized Lie case. 
On the one hand, the spacetime version just has the spacetime realization of everything.  
On the other hand, the canonical version has known \cite{Dirac, Battelle-DeWitt, DiracObs, K93, AM13} and more specialized nomenclature, 
meriting inclusion in round brackets below. 


\n 0) {\bf Closure} Given some initial candidate set of generators $\sbcG$, we assess them under Lie brackets. 
First-class generators close under these while second-class ones do not \cite{XIV}.  
(Canonically, constraints $\sbcC$ \cite{Dirac} play the role of generators.)
The Generalized Lie Algorithm \cite{ALRoPoT, XIV} permits the six following types of equation to arise from the generators $\sbcG$ 
(or constraints $\sbcC$, in which case one has the Dirac Algorithm \cite{Dirac, HTBook, ABook}).  

\n i)   {\it New generators} $\sbcG^{\prime}$ (or constraints $\sbcC^{\prime}$) arising as integrabilities are reliably found thus. 

\n ii)  {\it Identities}:      equations reducing to $0 = 0$. 

\n iii) {\it Inconsistencies}: equations reducing to $0 = 1$; including these incorporates \cite{ALRoPoT, XIV} an insight of Dirac's \cite{Dirac}, 
now freed from its more restricted context of Poisson brackets algebras of classical constraints. 
The Generalized Lie Algorithm thereby gains the capacity to reject candidate theories' sets of generators.  

\n iv)  {\it Rebracketing} using `{\it Lie--Dirac brackets}' in the event of encountering {\it second-class objects} 
(generalizing use of Dirac brackets \cite{Dirac, HTBook} to eliminate second-class constraints).  

\n v)  `{\it Specifier equations}' are also possible in the presence of an appending process. 
(Dirac's \cite{Dirac} appending of constraints to Hamiltonians $H$ using Lagrange multipliers $\bLambda$; 
$H \longrightarrow H + \bLambda \cdot \sbcC$ is the archetypal of a such.)
These specify what forms a priori free appending variables take. 

\n vi) {\it Topological obstruction terms} \cite{Dirac, ABook} such as anomalies [though this Letter just proceeds locally].   

\n The Generalized Lie Algorithm {\it terminates} if \cite{ALRoPoT, XIV} it 0)   {\it hits an inconsistency}, 
                                                                         I)   {\it cascades to inconsistency}, 
         												                 II)  {\it cascades to triviality}, or 
												                         III) {\it arrives at an iteration producing no new objects} 
																			            while retaining some degrees of freedom.   
\n Successful candidate theories do III), producing Lie algebraic structures of generators $\nFrg$ (or of first-class constraints).   

\vspace{0.07in}

\n 1) {\bf Relationalism} involves physical theories' physically meaningless transformations, 
which can be incorporated by means of Lie derivatives $\pounds_{\u{X}}$.  
In the canonical case \cite{ABook, ALRoPoT}, Relationalism splits into a) Temporal and b) Configurational: spatial plus internal. 
Manifest a) involves working with \cite{ABook, ALRoPoT} 
changes of configuration $\d \biQ (= \pounds_{\d} \biQ)$ 

\n in place of velocities $\dot{\biQ} = \d \biQ/d t$ so as to stay free from primary-level time variables.   
One is then not only to form reparametrization-invariant actions [variants of \cite{BSW} for GR], 
but also to stay within a Principles of Dynamics that uses changes instead of velocities among its variables \cite{ABook, ALRoPoT}.  
b) involves correcting by \cite{ABook, ALRoPoT} the Lie derivatives along the physically irrelevant group $\lFrg$ 
acting on configuration space $\FrQ$'s changes $\d \bia$: $\d \biQ \m \longrightarrow \m \d \biQ - \pounds_{\d \sbia} \biQ$.  
These corrections can be obtained systematically by solving geometrical level of structure $\bsigma$'s 
{\sl generalized Killing equation} $\pounds_{\u{X}} \bsigma = 0$ for the corresponding physically irrelevant automorphism group, $\lFrg$. 
a)              {\it provides} a primary constraint  quadratic in the momenta such as the GR Hamiltonian constraint $\scH$,   to be viewed as an equation of time;  
via this, time is to be abstracted from change.  
In contrast, b) {\it provides} [by variation with respect to $\bia$]  constraints linear    in the momenta such as     GR's momentum  constraint $\u{\scM}$.
In the spacetime case, there is a single Spacetime Relationalism.  
Where necessary, spacetime objects $\biS$ are now to be corrected  
by Lie derivatives along the physically irrelevant group $\lFrg_{\sS}$'s auxiliary variables $\biA$: $\biS \longrightarrow \biS - \pounds_{\sbiA} \biS$, 
which now provides generators.   
Relationalism can moreover also be viewed, in reverse [Fig 1.c)'s leftwards arrow], 
as {\it encoding} what [further] generators a theory needs for consistency. 

\vspace{0.07in}

\n 2) {\bf Observables} Given phase space or the space of spacetimes, observables $\sbiO$ are the {\it associated functions thereover}. 
In the presence of generators (or constraints), restricted (constrained) observables are such functions that additionally brackets-commute 
\cite{DiracObs, K93, AObs, ABook} with these.  
I.e.\ $\mbox{\bf [} \, \sbcG, \, \sbiO \, \mbox{\bf ]}  = 0$ or $\approx 0$: Dirac's notion \cite{Dirac} of weak equality extended to Lie Theory. 
The Jacobi identity moreover dictates \cite{AObs} that these notions only make sense after Closure has been ascertained (hence the downward arrow in Fig 1.c), 
and that these observables themselves close to form algebraic structures.  
Our zero brackets condition moreover translates to a first-order flow PDE system \cite{Lee2} 
amenable to (a slight extension of) Lie's Integral Approach to Geometrical Invariants \cite{Lie1890, Olver}.  

\vspace{0.07in}

\n 3) {\bf Constructability} One's algebraic structure of generators $\Frg$ is now to be {\it Deformed},  
$\bscG \m \longrightarrow \m \bscG_{\alpha}  \es  \bscG + \alpha \, \bphi$  for parameter $\alpha$ and functions $\bphi$, 
so as to see which members of the corresponding family of theories are also consistent.  
This uses the Generalized Lie Algorithm more extensively than Closure itself.  
This can be used to see if assumptions of less structure still suffice to give the same physical theory.  
The mathematical reason that this sometimes works is {\it Rigidity} of the underlying undeformed constraint algebraic structure; in particular GR is Rigid. 
This is in turn underlied by cohomological conditions $\mH^2(\Frg, \, \Frg) = 0$.   
Such Rigidity can moreover be taken \cite{XIV, UBIC} to provide a {\sl selection principle} in the Comparative Theory of Background Independence. 
Our algorithm can moreover {\it Bifurcate} \cite{HTBook}, 
corresponding to setting each of a string of multiplicative factors to zero giving a distinct consistent possibility. 
On the one hand, the Dirac Algorithm subcase of inconsistencies arising under Deformation is better known; see \cite{AM13, A-Brackets} and references therein. 
On the other hand, the Lie case's Deformations and Rigidities -- if not assessment of inconsistencies -- was done much earlier in the literature \cite{Def}.  
That the Generalized Lie Algorithm has the capacity to pull this off in cases other than the Dirac Algorithm 
is exemplified by provision of new foundations for Flat Geometry \cite{A-Brackets, XIV}. 
Namely that the two alternative top geometries here -- Conformal versus Projective -- arise as a Bifurcation from a Deformation and Rigidity analysis.  
This occurs both for Space from Less Space Structure assumed and for its indefinite flat spacetime counterpart.  

\vspace{0.07in}

\n On the one hand, each of 0) to 3) is at least somewhat well-known among theoretical physicists. 
On the other hand, putting these together as a coherent whole, is new, 
with the above overlaps forming Fig 1.c)'s `Lie claw' order of approach to the Problem of Time facets. 

\vspace{-0.15in}

\section{The Wheelerian two-way route}

The Wheelerian \c{MTW} two-way route between canonical and spacetime copies of the Lie claw consists of the following. 

\vspace{0.07in}

\n A) Another {\bf Constructability}: now {\bf of Spacetime from Assuming just Space} (within the remit of the Dirac Algorithm). 
Like all Constructabilities in this Article, this works by Deformation and Rigidity. 
In the case of GR, this gives the following \cite{AM13}. 

\n a) Einstein's dilemma of Galilean versus Lorentzian local Relativity now occurs as a Bifurcation of the Dirac Algorithm.
This illustrates inclusion of contracted limits \cite{Gilmore}.  

\n b) The relative coefficients within DeWitt's supermetric on GR's configuration space are derived.    

\vspace{0.07in}

\n B) In GR, {\bf Foliation Independence} is ascertained by {\bf Refoliation Invariance}, as illustrated in Fig 1.d-e).   
Teitelboim \cite{T73} showed this to be an implication of the Dirac algebroid formed by GR's constraints $\scH$, $\u{\scM}$.     

\vspace{-0.15in}

\section{Conclusion}

The overall order of approaching Problem of Time Facets, or equivalently of incorporating Background Independence aspects, is laid out in Fig 1.f). 
{\sl This answers the facet-ordering problem}.

The mathematics governing current Article's central the Lie claw digraph is moreover universal: {\it categorically meaningful}.
This is by the Lie claw's centre -- Closure -- being based on automorphism groups: a notion that is well-defined for every level of mathematical structure. 
Both this, and its overlaps with the other three facets of the problem, persist under both change of theory and passage to the quantum.   
This further significant point is covered further in \cite{UBIC}. 
Lie Theory has furthermore been substantially generalized since its inception in the late 19th century \cite{Lie1890}.
This means that, having spotted the mathematical nature of the Author's classical Local Resolution of the Problem of Time, 
over a century of further mathematical development can almost immediately be applied.
This further greatly clarifies the Nature of Physical Law at the Background Independent level.  
In fact, it is powerful enough that further Foundations of Geometry have dropped out of this work; Sec 3 gave an example; see also \cite{A-Brackets, XIV}.

Unlike Closure, Relationalism, Observables and Constructability, Foliation Independence is not a categorical name, 
so we introduce Intermediary-Object Independence.  
The putative strategy for resolving this is then not Refoliation Invariance but Reallocation of Intermediary-Object (RIO) Invariance. 
This retains the algebraic commuting-pentagon structure visible in Fig 1.e).  
I.e.\ whether going from an initial object to a final object, proceeding via intermediary object R or P 
causes one to be out by at most just an automorphism of the final object, $O_{21}^{\sf\si\sn} - O_{12}^{\sf\si\sn} = Aut(O^{\sf\si\sn})$. 
Both Constructability of Spacetime from Space and RIO Invariance are to be {\sl selection principles} 
in the Comparative Theory of Background Independence \cite{A-CBI, UBIC}.    

\vspace{0.07in}

\n{\bf Acknowledgments} I thank Professor Chris Isham for discussions over the past decade.
Professors Malcolm MacCallum, Don Page, Reza Tavakol and Enrique Alvarez for career support, and various close people for support.    

\vspace{-0.15in}


\end{document}